\documentclass[lettersize,journal]{IEEEtran}
\usepackage{siunitx}
\DeclareSIUnit\ohm{\ensuremath\Omega}
\DeclareSIUnit\db{dB}
\usepackage{cite}
\usepackage{amsmath,amssymb,amsfonts}
\usepackage{algorithmic}
\usepackage{graphicx}
\usepackage{textcomp}
\usepackage{xcolor}
\usepackage{url}
\usepackage[hidelinks]{hyperref}
\usepackage[nolist,nohyperlinks]{acronym}
\usepackage{footmisc}
\usepackage{caption}
\usepackage{subcaption}
\usepackage[inline]{enumitem}

\begin{acronym}
    \acro{STS}{Scrambled Timestamp Sequences}
    \acro{CIR}{Channel Impulse Response}
    \acro{UWB}{Ultra-Wideband}
    \acro{IC}{Integrated Circuit}
    \acro{IoT}{Internet of Things}
    \acro{I/O}{input/output}
    \acrodefplural{I/O}{inputs/outputs}
    \acro{MCU}{Microcontroller Unit}
    \acro{ToA}{Time of Arrival}
    \acro{TDoA}{Time Difference of Arrival}
    \acro{PDoA}{Phase Difference of Arrival}
    \acro{AoA}{Angle of Arrival}
    \acro{SPI}{Serial Peripheral Inteface}
    \acro{TWR}{Two-Way Ranging}
    \acro{GPS}{Global Positioning System}
    \acro{GNSS}{Global Navigation Satellite System}
    \acro{RTL}{Real-Time Localization}
    \acro{ToF}{Time of Flight}
    \acro{2D}{two-dimensional}
    \acro{3D}{three-dimensional}
    \acro{BOM}{Bill of Material}
    \acro{PCB}{Printed Circuit Board}
    \acro{LOS}{Line-of-Sight}
    \acro{NLOS}{Non-Line-of-Sight}
    \acro{BLE}{Bluetooth Low Energy}
    \acro{MSE}{Mean Squared Error}
    \acro{MAE}{Mean Absolute Error}
    \acro{ML}{Machine-Learning}
    \acro{CNN}{Convolutional Neural Network}
    \acro{NN}{Neural Network}
    \acro{SVM}{Support Vector Machine}
    \acro{FEM}{Finite Element Method}
    \acro{SMD}{Surface Mount Technology}
    \acro{PoA}{Phase of Arrival}
    \acro{RToF}{Return Time of Flight}
    \acro{RMS}{Root-Mean-Squared}
    \acro{M2M}{Machine-to-Machine}
    \acro{RF}{Radio Frequency}
    \acro{SDR}{Software-Defined Radio}
    \acro{DC}{direct current}
\end{acronym}

\def\BibTeX{{\rm B\kern-.05em{\sc i\kern-.025em b}\kern-.08em
    T\kern-.1667em\lower.7ex\hbox{E}\kern-.125emX}}

\usepackage{eso-pic}
\usepackage{url}
\AddToShipoutPictureBG*{
  \AtPageUpperLeft{%
    \put(0,-30){\raisebox{15pt}{\makebox[\paperwidth]{\begin{minipage}{21cm}\centering
      \textcolor{gray}{This article has been accepted for publication in the IEEE Transactions on Instrumentation and Measurement journal.\\DOI: 10.1109/TIM.2023.3282289} 
    \end{minipage}}}}%
  }
  \AtPageLowerLeft{%
    \raisebox{25pt}{\makebox[\paperwidth]{\begin{minipage}{21cm}\centering
      \textcolor{gray}{ \copyright 2023 IEEE.  Personal use of this material is permitted.  Permission from IEEE must be obtained for all other uses, in any current or future media, including reprinting/republishing this material for advertising or promotional purposes, creating new collective works, for resale or redistribution to servers or lists, or reuse of any copyrighted component of this work in other works.
      }
    \end{minipage}}}%
  }
}

\begin{document}
\bstctlcite{IEEEexample:BSTcontrol}

\title{Angle of Arrival and Centimeter Distance Estimation on a Smart \acs{UWB} Sensor Node}

\author{Tobias Margiani, Silvano Cortesi,~\IEEEmembership{Student Member}, Milena Keller, Christian Vogt,~\IEEEmembership{Member}, IEEE,\\ 
 Tommaso Polonelli,~\IEEEmembership{Member}, IEEE, Michele Magno,~\IEEEmembership{Senior Member}, IEEE
\thanks{Tobias Margiani, Silvano Cortesi, Milena Keller, Christian Vogt, Tommaso Polonelli and Michele Magno are with the D-ITET Department, ETH Zürich, Zürich, Switzerland (e-mail:
\{tobiasm, cortesis, milkelle, vogtch, topolonelli, magnom\}@ethz.ch).}
}

\markboth{IEEE TRANSACTIONS ON INSTRUMENTATION AND MEASUREMENT, Extension of SAS 2022 Manuscript No. 1570797111}%
{T. Margiani \MakeLowercase{\textit{et al.}}}




\maketitle

\begin{abstract}
  Accurate and low-power  indoor localization is becoming more and more of a necessity to empower novel consumer and industrial applications. In this field, the most promising technology is based on \ac{UWB} modulation; however, current \ac{UWB} positioning systems do not reach centimeter accuracy in general deployments due to multi-path and non-isotropic antennas, still necessitating several fixed anchors to estimate an object's position in space. This paper presents an in-depth study and assessment of \ac{AoA} \ac{UWB} measurements using a compact, low-power solution integrating a novel commercial module with \ac{PDoA} estimation as integrated feature. Results demonstrate the possibility of reaching centimeter distance precision and \ang{2.4} average angular accuracy in many operative conditions, e.g., in a \ang{90} range around the center. Moreover, integrating the channel impulse response, the phase differential of arrival, and the point-to-point distance, an error correction model is discussed to compensate for reflections, multi-paths, and front-back ambiguity. 
\end{abstract}

\acresetall
\section{Introduction}
\label{sec:introduction}
Today, precise and low-power indoor localization of mobile devices is still an open problem for both consumer and \ac{M2M} applications~\cite{cerro2022uwb}. Context and environmental awareness have been recognized as fundamental properties~\cite{zafari2019survey} in the \ac{IoT} ecosystem, not only by sensing the surroundings but also by estimating the spatial position regarding reference points and other \ac{IoT} agents~\cite{zhang2022demand}. In many application scenarios, relative and absolute location information are essential features, such as robotics, smart cities and smart manufacturing~\cite{zafari2019survey,chen2017robustness}. 
Currently, one of the challenges for indoor positioning is to find a sufficiently accurate indoor location method and, thus, a fitting technology~\cite{gifford2020impact} valid for extended areas, robust to changes to environmental conditions, scalable to support thousands of devices~\cite{zhao2021uloc}, and, at the same time, capable of working on battery supplied devices~\cite{polonelli2020flexible}. 
The usage of radio waves for indoor localization has proven to be one of the most solid methodologies, especially in the IoT ecosystem, where the same interface enables data transfer and localization~\cite{polonelli2020flexible}. In this context, \ac{UWB} is one of the most used pivotal technologies to overcome challenges related to the \ac{RTL} of objects in \ac{GPS}-denied areas. \ac{UWB} promises localization with centimeter, or even sub-centimeter~\cite{pala2021leading} accuracy, while consuming milliwatts of power and being computationally lightweight~\cite{polonelli2020flexible} - two prerequisites for \ac{IoT} devices. So far, \ac{UWB} could only provide a scalar distance \(l\) between two ranging agents, ignoring the direction of the incoming signals. However, recent advances~\cite{ledergerber2019angle,aoa_dw1000} demonstrate the possibility of extracting vectorial information \(\vec{l}\), thus providing two characteristics, distance \(l\) and direction \(\psi\), to the agent used as a reference point for estimating the spatial position. A vectorial point-to-point distance \(\vec{l}\) leads to an improved localization accuracy or/and allows for a decreased number of reference points to localize an object in \ac{3D} space~\cite{polonelli2022performance,ledergerber2019angle}.
In this direction, multiple approaches based on \ac{UWB} technology have been explored. Namely, \ac{AoA}, \ac{PDoA}, \ac{ToF}, \ac{TDoA}, and \ac{RToF} are techniques that show improved localization performance\cite{barbieri2021uwbaugmentation}. However, they often require an array of antennas, limiting the compactness of the designed solution~\cite{obeidat2021review}.
Many different approaches have been proposed to define a global standard similar to common outdoor technologies, such as \ac{GPS} and \ac{GNSS}~\cite{zhang2022demand}, exploiting other technologies such as WiFi, \ac{BLE}, \ac{UWB}, inertial sensors and others~\cite{cerro2022uwb, you2021ble, liu2021rfloc}. However, today there are no ready-to-use solutions for indoor localization that have been proven to be low-power, scalable, miniaturized, cheap, and accurate (centimeter precision), leaving this challenging research topic unsolved~\cite{cerro2022uwb}.

Among different \ac{IC} manufacturers, such as NXP\footnote{\scriptsize \url{www.nxp.com/applications/enabling-technologies/connectivity/ultra-wideband-uwb}}, SPARK Microsystem\footnote{\scriptsize \url{www.sparkmicro.com}}, Microchip\footnote{\scriptsize \url{www.microchip.com/en-us/products/wireless-connectivity/ultra-wideband-solutions}}, and Qorvo\footnote{\scriptsize \url{www.qorvo.com}}, products of the last are some of the most widely used \ac{UWB} modules, namely Decawave's DWM1000 family~\cite{mayer2019embeduwb,polonelli2020flexible,pala2021leading}. It is a commercial \ac{UWB} \ac{IC} that supports \ac{ToF} measurements to estimate the distance among generic devices. Moreover, by estimating the distances between one node and at least four anchors, a trilateration algorithm can be applied to estimate its location in \ac{2D} and \ac{3D} space. 
However, recent studies~\cite{polonelli2020flexible, mayer2019embeduwb,polonelli2022performance} have shown that the ranging accuracy is still not reliable in several conditions (i.e., \ac{NLOS} or presence of external electromagnetic noise).
The recently released DW3xxx family from Qorvo showed an increased accuracy paired with a two-times  reduction in power consumption \cite{polonelli2022performance}. In detail, the DW3120 and DW3220 \ac{UWB} transceivers support two-way ranging, \ac{TDoA} and \ac{PDoA} implementations, enabling not only the estimation of a scalar distance but the possibility to measure at the same time the received signal's \ac{AoA} and \ac{ToF}. Thus, providing a \ac{2D} position estimation (\(\vec{l}\)) relying on one anchor only. The only alternative \ac{UWB} module that offers similar \ac{AoA} capabilities within a single \ac{IC} is the NXP SR150. In contrast to the Qorvo DW3x20's claimed accuracy of \(\pm\ang{5}\) at \(\pm\ang{45}\)~\cite{Qorvo_DW3220}, the NXP SR150 has a maximum error of more than \(\ang{16}\) at \(\ang{-45}\)~\cite{NXP_AoA_Paper}. Other options for achieving similar accuracies as the DW3x20 are using multiple UWB modules (e.g. two DW1000) and therefore doubling the power consumption. 
\par
Optimized for low-power battery-supplied operation, the DW3xxx family is designed to be used in mobile, consumer, and industrial applications~\cite{stocker2022uwb, polonelli2022performance}. However, at the moment of this paper's writing, no existing investigations or extensive studies demonstrate its effective performance in the field. 
Extending on the previous study on UWB localization~\cite{polonelli2022performance}, this work presents a sensor node designed to enable vectorial distance estimation (in \ac{2D}) on a compact, low-power, and plug-and-play solution for many different applications, such as indoor localization for robotics, wearable, and \ac{IoT}. Results show that the \ac{AoA} accuracy reaches an average of \ang{0.59} in the \(\pm\)\ang{45} around the center with a compact \qtyproduct{3 x 3.5}{\centi\meter} \ac{PCB} and an average power consumption of \qty{55}{\milli\watt} while in ranging mode. The sensor node is characterized and assessed through a set of data collected in real operational environments, discussing its strengths and weakness, such as front-back ambiguity around \ang{180} \ac{AoA}. Different error compensation methodologies are proposed, demonstrating the possibility of correcting the measurement uncertainty with a non-linear regression model (up to \(35\times\) improvement in accuracy) or classifying reliable operative zones to invalidate values collected outside the confidence interval (\(>\)90\% classification accuracy). In detail, the scientific content of this paper includes: 
\begin{enumerate*}[label=(\roman*),,font=\itshape]
  \item Theoretical and background study to enable the \ac{AoA} estimation over the standard \ac{UWB} \ac{TWR} protocol; 
  \item Design and implementation of a compact sensor node for vectorial distance measurements \(\vec{l}\) exploiting the Qorvo DW3220 in a dual antenna setup;
  \item A dataset including several thousand points collected in real operative conditions, acquired at \ang{1} incremental \ac{AoA} steps for 11 distances in a point-to-point \ac{TWR} configuration;
  \item A system-level characterization reporting the measurement accuracy in each studied operational condition, including observed boundaries, measurement stability, and repeatability; 
  \item Based on the physical characterization, error correction methodologies are discussed, demonstrating the possibility of compensating for non-linear system behavior through regression and classification models. 
\end{enumerate*}
This work can drive future researchers and engineers to enhance indoor localization systems exploiting 2D vectorial distance measurements \(\vec{l}\) in miniaturized and low-power devices, providing a system description and also discussing challenges and limitations to enable accurate \ac{UWB}-\ac{AoA} measurements in the field.  Lastly, the \ac{PCB} design and the dataset are released on GitHub as open-source files\footnote{\label{txt:github}\scriptsize\url{github.com/ETH-PBL/UWB_DualAntenna_AoA}}.

\section{Related Works}
    The physical principles of \ac{AoA} estimation based on radio waves have been known for over 80 years~\cite{friis_1934_aoa}. So far, indoor localization techniques using a single receiver were often based on \ac{SDR} and for the use-case optimally tuned carrier frequencies and bandwidths~\cite{li2021decimeter, neunteufel2022narrowband}. In recent years, the availability of low-cost hardware and the need for a reliable and cheap indoor localization system pushed the development of novel technologies~\cite{barbieri2021uwbaugmentation}. In 2015, Dotlic \textit{et al.}~\cite{aoa_dw1000} described \ac{AoA} estimation using \ac{UWB} with a dual-transceiver board, achieving an estimation error of \(<\)\ang{10} for angles in \(\pm\ang{80}\). Starting with the introduction of \ac{BLE} Direction Finding in 2019, the market started to push towards a low-cost and widely available technology for \ac{AoA} estimation using a single transceiver. In the work of Botler \textit{et al.}~\cite{uwb_ble_comparison}, \ac{AoA} estimation using \ac{UWB} is compared against \ac{BLE}. Although the physical principle for determining the \acl{AoA} is the same for both \ac{UWB} and \ac{BLE}, \ac{UWB}-\ac{AoA} suffers less from an obstructed \ac{LOS} and Rayleigh fading due to the higher bandwidth~\cite{bott2021uwb, aoa_dw1000}. Measurements performed by Botler et al. from \(\ang{\pm90}\) at a \qty{3}{\meter} distance between anchor and tag showed that \ac{UWB} generally achieved an accuracy of up to \ang{5}, despite obstructed \ac{LOS} and multi-path propagation. In comparison, \ac{BLE} achieved an average error of almost \ang{25} in an outdoor scenario without obstacles. This shows that for an accurate estimation of the \ac{AoA}, \ac{UWB} is the most promising technology to use. In order to minimize the influence of multi-path propagation, we therefore decided to focus on a \ac{UWB} only \ac{AoA} estimation.
    \par
    To further decrease the error from the \ac{UWB} dual-transceiver approach, originating in scheduling uncertainties of the DW1000, \cite{heydariaan_2020} introduced a way to synchronize the two transceivers by reducing the timing jitter. Using an additional timestamp correction, the localization accuracy increased by up to 44.33\%. The proposed technique has been tested against five responders concurrently. The achieved \ac{AoA} error is below \ang{10} in 90\% of the measurements. The DW3220 used in our paper adds support for a second antenna and on-chip \ac{TDoA} and \ac{PDoA} estimations, thus reducing the timing jitter introduced by the scheduling on two separate transceivers.
    \par
    In addition to the timing jitter, the indoor environments introduced a significant source of estimation errors (both for distance and angle estimation). Objects within the environment influence wave propagation and cause absorption, diffraction, and multi-path scattering~\cite{molisch2009channels}. In~\cite{stahlke2020nlos}, authors applied different \ac{ML} techniques (\acp{CNN} and \acp{SVM}) to detect \ac{NLOS} conditions directly from \ac{CIR} measurements and compensate for it in ranging applications. Using six receivers, their evaluation took place as the first step in a known environment (corridor). The different deep neural networks achieved a detection accuracy of over 94\%, leaving the \ac{SVM} approaches behind by over 10\%. Second, they tried it within a generalized, unknown environment. Also there, the \ac{CNN} outperformed the \ac{SVM} by 10\%, reaching an accuracy of \(>\)90\%. In a last industrial setting, they tried to augment the localization accuracy and enhanced the \ac{MAE} of ranging from \qty{17}{\centi\meter} to \qty{12}{\centi\meter}. In~\cite{ledergerber2019angle}, the \ac{AoA} is estimated using a single DW1000. Due to the non-idealities of the antenna, the \ac{CIR} varies between different \acp{AoA}. By performing multiple measurements at the same location but with different antenna orientations, Ledergerber \textit{et al.} could map the \ac{CIR} to the \ac{AoA}. The \ac{CIR} is measured at different angles in \(\pm\ang{180}\) and then classified into a maximum of 256 bins using a \ac{NN} representing the different angles (resolution of \ang{1.41}). With ten consecutive measurements (\(N_{\mathrm{CIR}}=10\)), 90\% of the predictions had an absolute error of less than \ang{25}. The predictor is then applied to localize a mobile robot, using multiple transmitters and a particle filter to track the robot's position estimation. The overall achieved position estimate had a \ac{RMS} error of \qty{0.37}{\meter} and \ang{3.6}. Instead of classifying the \ac{CIR} into different bins, \cite{naseri2022machine} applied a deep \ac{CNN} to perform a regression-based \ac{AoA} estimation using an \(N_{\mathrm{CIR}}\) of 5. The \ac{ML}-based solution outperforms the classical \ac{PDoA} approach in the range of \(\pm\ang{90}\), mainly due to the non-linearity \ac{PDoA} suffers from around extreme angles. In an experimental setup, including a noisy and multi-path environment, the proposed solution has a 99\textsuperscript{th} percentile error of \ang{2.8}, compared to the \ang{84.5} of \ac{PDoA}. In this paper, we combine these two \ac{ML} approaches to compensate for non-idealities, by first splitting the angles into bins/zones and then applying error corrections optimized for each specific zone.

\section{\acs{AoA} Estimation}
\label{sec:aoa-and-distance-estimation}

The distance between two \ac{UWB} transceivers is estimated based on the \ac{ToF} of the wireless signal. 

%

\begin{figure}[t]
\centering
\centerline{\includegraphics{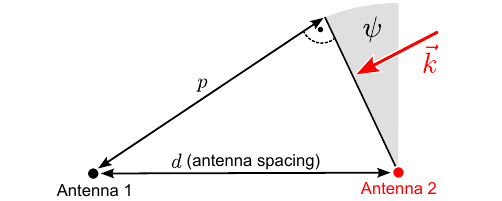}}
\caption{Dual antenna setup with distance between the antennas \(d\), actual angle of arrival \(\psi\), incident plane wave (vector \(\vec{k}\)) and resulting antenna signal phase difference caused by additional travel distance \(p\).}
\label{fig:aoa}
\end{figure}

In addition to the distance measurement, the DW3220 \ac{IC} can measure \ac{TDoA} and \ac{PDoA}. The \ac{TDoA} indicates the timestamp difference of the same signal arriving at the two antennas. Since the typical antenna spacing (\(d\) in \autoref{fig:aoa}) is half a wavelength or closer, a typical \ac{TDoA} value is below \qty{1}{ns}. On the other side, the \ac{PDoA} measures the difference in signal phase (due to additional travel distance \(p\) in \autoref{fig:aoa}) between the first antenna and the same signal arriving at the second antenna. With appropriate antenna placement, the \ac{AoA} (\(\psi\)) is strictly correlated with the \ac{PDoA}.
Thus, the planar (\ac{2D}) \ac{AoA} can be calculated from the \ac{PDoA} \(\alpha\), using a geometrical approach shown in \autoref{fig:aoa} and expressed in \autoref{eq:aoa}, with the carrier wavelength defined as \(\lambda\).

\begin{equation}
  \psi = \arcsin \frac{\alpha\lambda}{2\pi d}  
  \label{eq:aoa}
\end{equation}

There are two points to note when using a dual antenna system to estimate the \ac{AoA}. First, by using only the \ac{PDoA} measurement, it is impossible to differentiate between signals arriving from the ``front'' of the module and those arriving from the ``back''. This effect is called front-back ambiguity~\cite{heydariaan_2020}, and limits the absolute angle measurement to \ac{2D} and \(\pm\)\qty{90}{\degree}. Second, for accurate localization, the distance and \ac{AoA} measurements must result from signals traveling on a \ac{LOS} path. Otherwise, the incoming signal can result from a multi-path, thus arriving from a completely different direction. Hence, it becomes clear that a compensation algorithm, or a system with \(>2\) antennas, can be beneficial to correct the aforementioned source of inaccuracies, also enlarging system boundaries to cover the entire \qty{360}{\degree} range.


\section{Hardware Description}
\label{sec:hardware-description}
The hardware used in the proposed system is split into three main
components. First, an STM32-based \ac{MCU} board from STMicroelectronics (later referred to as host \ac{MCU}) interfacing with the \ac{UWB} transceiver and a computer for data collection. Second, a custom \ac{PCB} containing a Qorvo DW3220 \ac{UWB} transceiver connected to two antennas (later referred to as double antenna module). Lastly, a Qorvo DWM3000EVB board with a Qorvo DW3110 transceiver and a single antenna as a communication counterpart to the custom dual antenna module (later referred to as single antenna module) was used alternatively as an anchor or tag depending on the specific test setup.
The choice of an STM32 \ac{MCU} allows for rapid development using the Qorvo \ac{UWB} \ac{IC} and easy integration into future low-power localization applications.

\subsection{DW3220 \acs{UWB} Module}
\label{sec:dw3220}

The Decawave/Qorvo DW3220\footnote{\url{www.qorvo.com/products/p/DW3220}\label{fnlabel}} \ac{IC} is a low-power IEEE 802.15.4-2015 compatible \ac{UWB} transceiver released in 2020. It is the successor of the widely used Decawave DW1000 \ac{IC} \cite{coppens2022overview}, adding support for a second antenna and on-chip \ac{TDoA} and \ac{PDoA} estimations. The claimed typical accuracy is \(\pm\qty{6}{\centi\meter}\) for ranging and \(\pm\ang{10}\) for PDoA\footref{fnlabel} estimation in \ac{LOS} conditions. While there has been a range of research on using multiple antennas to improve localization~\cite{naseri2022machine}, the DW3220 \ac{IC} is, to the best of our knowledge, the first supporting two antennas and \ac{PDoA} estimation in a single compact package and optimized power consumption. The DW3000 series of \acp{IC} supports \ac{UWB} channels 5 and 9, data rates of \SI{850}{\kilo bps} and \SI{6.8}{\mega bps}, \ac{STS} for secure ranging, and multiple low-power states. The DW3220 has two antenna ports internally connected to a \ac{RF} switch and a single radio front end. The switch is automatically controlled in hardware to sample the \ac{CIR} data from both antennas and measure \ac{TDoA} and \ac{PDoA}. This automatic switch control is only possible if the incoming data packet contains an \ac{STS} frame, a feature added with IEEE 802.15.4z, which therefore is not compatible with the older DW1000 \ac{IC}. A single \ac{IC} with two antenna ports eliminates the need for pico-second time synchronization between multiple receivers, which would otherwise be required for \ac{TDoA} and \ac{PDoA} estimation. Additionally, it reduces the \ac{BOM}, simplifies dual antenna designs, optimizes power consumption (e.g., \(4\times\) reduction compared to \cite{naseri2022machine}), and results in a potentially higher accuracy by eliminating timing error sources.

\subsection{Sensor Node}\label{sec:custom_pcb}
At the moment of this paper's writing, no evaluation boards, commercial products or hardware design documentation is publicly released. Thus, to use the DW3220 \ac{IC}, a sensor node was designed to support the double antenna configuration and satisfy \ac{RF} minimal requirements. It is optimized for low power usage and \ac{RF} quality when \ac{UWB} channel 5 (\qty{6.5}{\giga\hertz}) is used. Two chip antennas are spaced at \(1/2\) wavelength to allow accurate \ac{PDoA} estimation - a design specification from a previous study \cite{aoa_dw1000}.

The power supply is divided into two voltage regulators with a dedicated \ac{DC}/\ac{DC} converter each. The external power supply is connected to the module through a castellated connector. It is linked to the first buck converter, TPS62233DRYR, which generates an output voltage of \SI{3}{\volt}. This feeds the DW3220 VDD1 main power supply, which supplies the device \acp{I/O} and the always-on domain. The \SI{3}{V} line powers the second buck converter TPS62743YFPR with a fixed \ac{DC} output voltage of \SI{1.8}{\volt}, connected to the VDD3 pin. The TPS62743YFPR is controlled by the DW3220, which can enable or disable the \ac{RF} power supply in real time - depending on the operational mode. Using this scheme, the most efficient power consumption setting is achieved.

\begin{figure}[t]
\centerline{\includegraphics{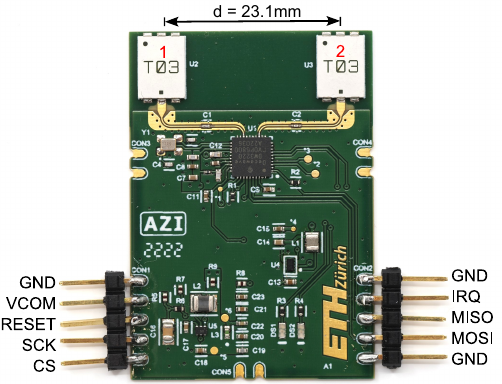}}
\caption{Sensor node. Antenna intra-spacing and digital connections are outlined.}
\label{fig:custompcb}
\end{figure}

As explained in Section~\ref{sec:aoa-and-distance-estimation}, the antennas must be separated by \(\lambda/2\) to reach the best accuracy. We use channel 5 with a nominal frequency of \(f = \SI{6.4896}{GHz}\) and therefore an antenna spacing of \(d = c/(2*f) = 0.023114 \approx \SI{2.31}{cm}\). The desired \SI{50}{\ohm} impedance target, the microstrip to coplanar waveguide coupling, and the choice of the dielectric material are investigated and simulated to match the DW3220 specs. An S11 of \SI{-22}{\db} was reached with a \textit{ThunderClad 2} substrate material around \SI{6.5}{\giga\hertz} and a bandwidth of \SI{4}{\giga\hertz} @ \(<\)\SI{-10}{\db}. The finally selected substrate was \textit{ThunderClad 2} with \SI{0.1068}{\milli\meter} thickness and a trace width of \SI{0.145}{\milli\meter}.

To correctly estimate the antenna's electrical delay and the cross-coupling between the two, together with the radiation pattern, a precise antenna \ac{3D} \ac{FEM} model is required. We selected the commercial integrated \ac{SMD} antenna AH086M555003 manufactured by Taiyo Yuden for our design. The \ac{3D} \ac{FEM} tools ANSYS HFSS\footnote{\url{https://www.ansys.com/products/electronics/ansys-hfss}} and ANSYS SIwave\footnote{\url{https://www.ansys.com/products/electronics/ansys-siwave}} were used to additionally simulate the behavior of the \ac{RF} traces to ensure proper matching between the chip-antennas and the controlled impedance \ac{RF} traces on the sensor node as well as with the addition of the real antenna model.

\begin{figure}[t]
    \centering
    \begin{subfigure}[t]{1\columnwidth}
        \centering
        \includegraphics[width=\columnwidth]{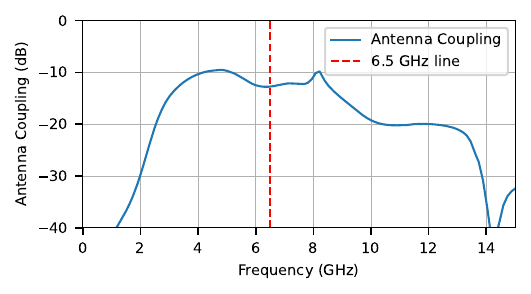}
        \caption{Coupling factor between the two AH086M555003 antennas.}
        \label{fig:coupling}
    \end{subfigure}
    \\
    \hfill
    \centering
    \begin{subfigure}[t]{1\columnwidth}
        \centering
        \includegraphics[width=\columnwidth]{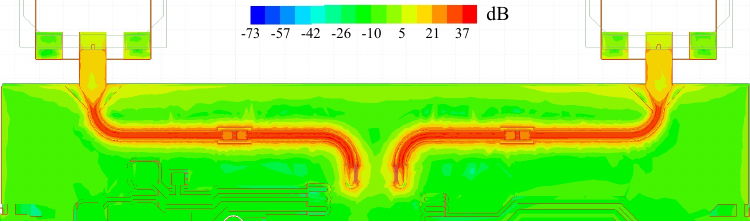}
        \caption{Current density distribution over the \ac{RF} traces, matched at \SI{50}{\ohm} and symmetrically placed.}
        \label{fig:currdens}
    \end{subfigure}
    \caption{FEM simulation and modeling of the designed dual antenna \acs{PCB}.  The \acs{PCB} substrate is \textit{ThunderClad 2} with \SI{0.1068}{\milli\meter} thickness between the \ac{RF} line and the ground plane. The two antennas are at \SI{2.31}{cm} distance.}
    \label{fig:cf_tof_tot}
\end{figure}

\autoref{fig:coupling} shows the graph of the antenna coupling parameter expressed in decibels. At a frequency of \SI{6.5}{\giga\hertz}, the coupling amounts to \SI{-12.55}{\db}. This parameter is not optimal but acceptable for the scope of this work. For future design and developments, a dedicated custom dual-antenna should be considered to decrease, or compensate, for this parasitic effect. \autoref{fig:currdens} shows the simulated current density distribution in the region of the \ac{RF} traces. As expected, the traces are conducting most of the energy, expressed by the red color scale corresponding to a high current density. The coupling of the two antennas is visible, as exactly in the middle between the two lines, the current density on the ground plane is not equal to zero and, therefore indicating a non-perfect decoupling.

\section{Empirical Characterization}
\label{sec:empirical-characterization}

To demonstrate the possibility of gathering \ac{AoA} and distance information using a compact integrated sensor, a range of tests were conducted to quantify the performance of the DW3220 \ac{IC} empirically and, at the same time, collecting a comprehensive dataset to model and compensate for errors and measurement inaccuracies. For all the presented measurements, one single antenna module and one double antenna module were placed in an \qtyproduct{8.6 x 7.7}{\meter} room containing tables and office equipment placed along the walls and a \qtyproduct{6.5 x 5.5}{\meter} empty space where the measurements were performed. The single antenna module was fixed on a tripod at a height of \qty{1.1}{\meter}. In contrast, the double antenna module was mounted on a rotatable motorized platform, providing a ground truth \ac{AoA} with a precision of \ang{1}. The modules were configured to perform two-way ranging \cite{polonelli2022performance} at a frequency of \qty{4.6}{\hertz}. All measurements were done with \ac{LOS} visibility between the two \ac{UWB} modules, and no objects were placed within \qty{1}{\meter} of the devices or the \ac{LOS} signal path.

The two \ac{UWB} modules continuously exchange \ac{TWR} frames while a local computer logs each frame received by the double antenna module. The collected data includes the \ac{CIR} from both antennas, \ac{TDoA}, \ac{PDoA}, transmission quality and signal strength indicators, as well as round-trip and reply times from \ac{TWR}, and the computed distance estimation. All data, except for the \ac{TWR} information, is directly computed on hardware, therefore, requires no processing on the local \ac{MCU}.

The dataset used for the system characterization and later error compensation (\autoref{sec:error-compensation}) contains measurements collected from a total of 19880 exchanges. For data collection, the \ac{UWB} modules were placed at 11 fixed distances ranging from \qty{50}{\centi\meter} to \qty{5.5}{\meter} in \qty{50}{\centi\meter} steps. At each distance, the double antenna module was rotated in \ang{1} steps, and for each of those, at least five \ac{TWR} measurements were performed.

\subsection{\acs{PDoA} and \acs{AoA}}
\label{sec:empirical-pdoa-and-aoa}

\begin{figure}[t]
    \centering
    \includegraphics[width=0.9\columnwidth]{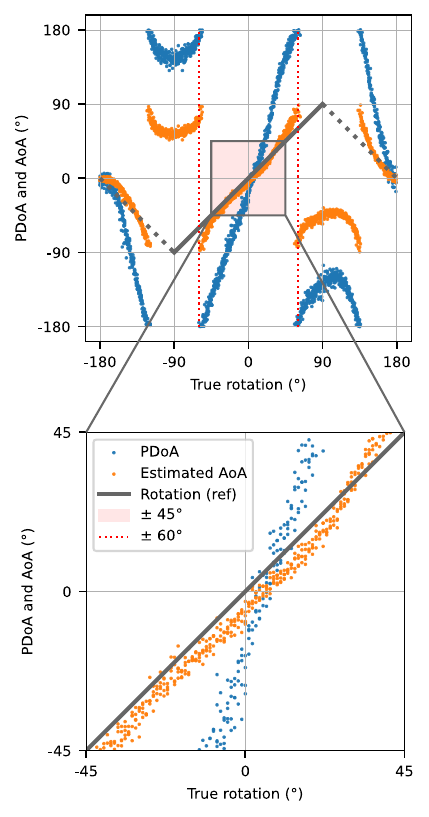}
    \caption{\acs{PDoA} measurement and resulting \acs{AoA} estimation (based on \autoref{eq:aoa}) at a distance of \qty{3}{\meter}. Note that the true rotation was a full \ang{360}; however, due to front-back ambiguity, all estimated \ac{AoA} values fall between \ang{-90} and \ang{90}. Outside of this range, the reference line is dotted, showing the expected \ac{AoA} estimate instead of continuing straight on the true \ac{AoA} values.}
    \label{fig:pdoa-and-aoa}
\end{figure}

From each collected measurement, an \ac{AoA} estimation was computed using \autoref{eq:aoa} and coupled with its ground truth angle. An example of the collected data is shown in \autoref{fig:pdoa-and-aoa}, where the device-to-device distance is \qty{3}{\meter}. \autoref{fig:pdoa-and-aoa} shows that the \ac{AoA} estimation curve is linear around \(\pm\)\ang{45} and strongly correlates (\(r = 0.98\)) with the ground truth on the horizontal axis. For angles around \(\pm \ang{60}\), a non-linear behavior is apparent, introducing an increase of the average error above \ang{5}. Outside this range, the AoA estimation accuracy drops significantly, losing a visible correlation with the actual rotation. Additionally, front-back ambiguity can be observed outside the center \(\pm\ang{90}\) range. The \ac{PDoA} and estimated \ac{AoA} values turn back towards \ang{0} instead of increasing towards \ang{180}. As this is expected from the physics of the double antenna setup and simulated in Section~\ref{sec:3DFEM}, the reference line is drawn accordingly (dashed in \autoref{fig:pdoa-and-aoa}).

However, despite non-linear behavior and the front-back ambiguity being expected around \ang{90}, results in \autoref{fig:pdoa-and-aoa} show a wide glitch of \ang{63} to \ang{77} starting at \(\pm\ang{60}\), equivalent to a variation of the antenna spacing \(\Delta d\) of \qty{7.7}{\milli\meter}. This wrap-around centered at \(\pm\)\ang{90} in the \ac{PDoA} could be caused by several possible factors. Among them, we considered an incorrect antenna placement (see \autoref{fig:custompcb}) during the \ac{PCB} assembly, which was disproved by observing similar results on two additionally soldered \acp{PCB}. A second alternative is the effective antenna center and signal propagation, which was considered an ideal point in the center of the \ac{SMD} package and, more important, spatially invariant. In reality, the antenna delay and its effective wave incident point can vary depending on the angle of the incoming signal, consequently boosting the front-back ambiguity error on a wider angular range.

Considering only true \ac{AoA} values from \ang{-45} to \ang{45} (as pictured in the zoom of \autoref{fig:pdoa-and-aoa}), the \ac{AoA} estimate reaches a \ac{MSE} of only \ang{0.59} and an \ac{MAE} of \ang{5.21} at a distance of \qty{3}{\meter}. For all evaluated distances, the maximum \ac{MSE} is \ang{2.40} and the maximum \ac{MAE} \ang{8.46}. Extending the range up to where the wrap-around in \ac{PDoA} values begins increases the error due to the denoted non-linear behavior. At a \qty{3}{\meter} distance and a range of \(\pm\ang{60}\), an \ac{MSE} of \ang{12.00} and \ac{MAE}
of \ang{11.67} is achieved. The precise angle of the wrap-around varies depending on the distance but always lies between \ang{50} and \ang{60}. Another observation based on the collected data is a flattening of the \ac{PDoA} and \ac{AoA} curves close to \ang{180} and increased non-linearity towards \(\pm \ang{135}\). This can be seen in \autoref{fig:pdoa-and-aoa} by comparing the shape around \ang{0} in the center and \ang{180} on the left and right sides of the plot. The distance from \ang{180} to the wrap-around in \ac{PDoA} values is only \ang{37} to \ang{47}, whereas the distance from \ang{0} to the wrap-around is \ang{50} to \ang{60}. These considerations demonstrate a decreased accuracy while operating on the back side of the module. In the \ang{90} range from \ang{135} to \ang{225}, an \ac{MSE} between \ang{13.51} and \ang{23.95} is observed depending on the distance.

An analysis of the raw data originating from the \ac{TWR} distance measurement will be skipped here, as there are only a few changes compared to the well-studied DW1000 \ac{IC}, which is already extensively discussed in \cite{polonelli2022performance}. The data will, however, be used in \autoref{sec:error-compensation} to develop a joint error correction model for \ac{AoA} and \ac{TWR} distance errors.
The collected dataset and the \ac{PCB} design are released as open source on GitHub for future usage and improvements\footref{txt:github}.

\subsection{Stability}
\label{sec:empirical-stability}

\begin{figure}[t]
\centerline{\includegraphics[width=\columnwidth]{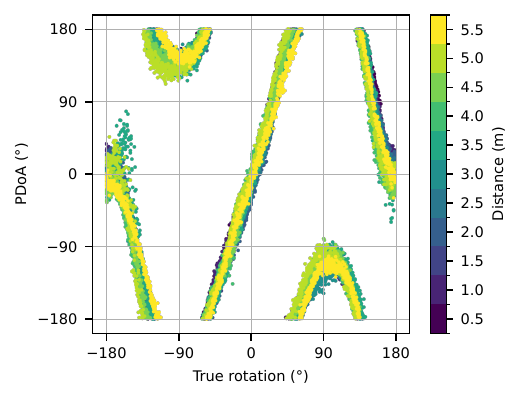}}
\caption{\acs{PDoA} measurements for each \ac{AoA} in \ang{1} steps and each distance between \qty{50}{\centi\meter} and \qty{5.5}{\meter} in \qty{50}{\centi\meter} steps.}
\label{fig:pdoa-distance}
\end{figure}

The stability of the \ac{AoA} estimation provided by the DW3220 is evaluated over different distances and time. As shown by the experimental results in \autoref{fig:pdoa-distance}, the \ac{PDoA} measurements are comparable over all tested distances, between \qty{50}{\centi\meter} and \qty{5.5}{\meter}. The \ac{MSE} and \ac{MAE} are consistent with the characterization mentioned above at \qty{3}{\meter} and within the DW3220 specs. Noticeable differences are the precise point of the wrap-around in \ac{PDoA} values, which is spread over \ang{25}, and the different inaccuracies around \ang{180}.
\par
To evaluate the stability over time, another set of experiments was performed. The two \ac{UWB} modules were again placed at a fixed distance, and it was ensured that there was no movement in the room for the duration of each experiment. Over a period of 11 minutes, about 3000 \ac{TWR} exchanges were performed. This was repeated for \ac{AoA} of \ang{0}, \ang{90}, \ang{180} and \ang{270}. The resulting \ac{PDoA} measurements feature no drift over time, with results spread as Gaussian distribution. After computing the estimated \ac{AoA}, a standard deviation of \ang{2.45} is achieved over all tested angles. By averaging 10 \ac{AoA} estimations, the standard deviation can be reduced to below \ang{1} for all angles, and in the best case (\(\psi=\ang{0}\)), even below \ang{0.6}. When averaging results, additional considerations are necessary to properly handle values close to the boundaries as, e.g., a true \ac{PDoA} value close to \ang{+180} can also be measured as close to \ang{-180}.

\subsection{\acs{3D} \acs{FEM} \acs{PCB} Simulation}
\label{sec:3DFEM}
In order to analyze some of the non-idealities encountered during the empirical characterization of the \ac{PCB} designed in Section~\ref{sec:custom_pcb}, a \ac{3D} \ac{FEM} simulation was conducted to compare the phase shift between the antennas with known angles of arrival. The \ac{PCB} design was imported into Ansys HFSS \ac{3D} together with a model of the \ac{UWB} antenna. The environment is a \SI{120}{\milli\meter}-sized vacuum cube with a perfectly matched boundary layer. The simulation is based on a transient composite excitation model with a plane wave source excitation. 
\begin{figure}[t]
\centerline{\includegraphics{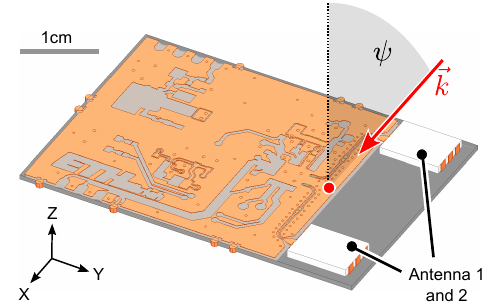}}
\caption{\acs{3D} rendering of the simulated \ac{PCB} geometry with the origin marked as a red dot between the two antennas and the wave incident angle \(\psi\), as well as an example of the plane wave propagation vector \(\vec{k}\).}
\label{fig:fem_setup}
\end{figure}

The plane wave's phase origin (\(\vec{O}\)) is rotated along a circle with a \SI{50}{\milli\meter} radius centered between the two antennas at the \ac{PCB} simulation origin (see \autoref{eqn:zerophase} and the center in \autoref{fig:fem_setup}). This was chosen as a compromise between the absolute distance of the wave source to the \ac{PCB} and relative distance difference to the antennas. The incident angle of the wave with respect to the \ac{PCB} plane normal is defined as \(\psi\).

\begin{equation}
    \vec{O} = \SI{50}{\milli\meter} \cdot \left(\begin{array}{c} \sin(\psi) \\ 0 \\ \cos(\psi) \end{array} \right)
    \label{eqn:zerophase}
\end{equation}

The wave propagates along a \(\vec{k}\)-vector pointing to this simulation origin (\autoref{eqn:kvect}), and the electric field vector points along the \(y\)-axis of the simulation. 
\begin{equation}
    \vec{k} = \left(\begin{array}{c} -\sin(\psi) \\ 0 \\ -\cos(\psi) \end{array} \right)
    \label{eqn:kvect}
\end{equation}

The time-dependent electric field magnitude \(A(t)\) (\autoref{eqn:uwb_pulse}) of this plane wave is defined to be a \ac{UWB} pulse with a bandwidth of \(\approx\)\SI{500}{\mega\hertz} (\(\tau\) = \SI{4}{\nano\second}), a center frequency of \(f_0\) = \SI{6.5}{\giga\hertz}, and a scale factor \(G\) based on \cite{ghavami_novel_nodate}. The \ac{UWB} receiver chip was modeled as lumped element resistors of \SI{50}{\ohm} located at the solder pads of the \ac{RF} input to the chip and referenced to the ground plane next to these pads.

\begin{equation}
    A(t) = G\cdot \frac{t}{\tau} \cdot e^{-2\pi\left(\frac{t}{\tau}\right)^2}\cdot \sin(2\pi f_0 t)
    \label{eqn:uwb_pulse}
\end{equation}

During the simulation, \(\psi\) varied between \SI{0}{\degree} and \SI{90}{\degree}, and the resulting phase shift between the antennas was extracted. The simulation had \(\approx 245200\) degrees of freedom to solve and a transient time between 0 and \SI{4}{\nano\second}. 

\begin{figure}[t]
\centerline{\includegraphics{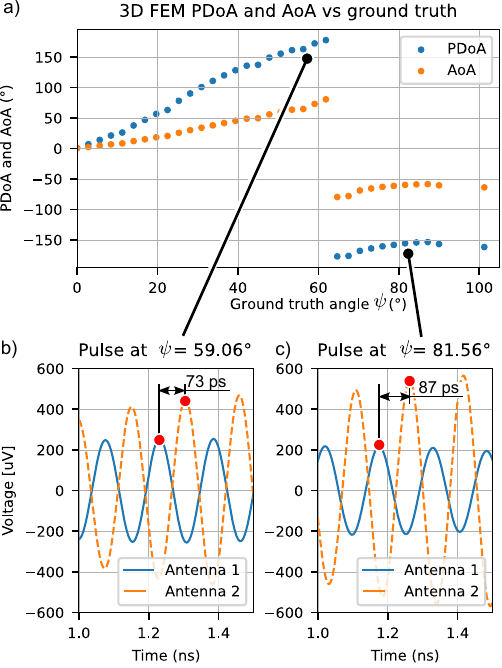}}
\caption{Resulting phase shift between antennas with respect to known input angle \(\psi\), with an overview of all \ac{PDoA} and corresponding \ac{AoA} calculated with \autoref{eq:aoa} (a), detail \ac{TDoA} for the wrap-around effect of angles (b,c).}
\label{fig:fem_result}
\end{figure}

The resulting phase shift with respect to the incident wave is given in \autoref{fig:fem_result}. It shows the \ac{PDoA} phase wrap-around at \(\psi>\approx\SI{65}{\degree}\) (\autoref{fig:fem_result}a). The insets in \autoref{fig:fem_result}b,c present the voltage received at the \ac{UWB} chip inputs, caused by the incident plane wave. It illustrates the phase-wraparound due to a too-large time (phase) shift between the two antennas. At \(\psi\) = \ang{81.56} (\autoref{fig:fem_result}c), the time shift between the antennas is \qty{87}{\pico\second}, corresponding to \(\approx\ang{203}\) at \(f_c=\qty{6.5}{\giga\hertz}\). Due to the possible solutions of \ac{PDoA}, this wraps around to \(\approx\ang{-157}\). Therefore, future PCB designs could be improved by reducing the antenna distance.

\section{Position and Angle Error Compensation}
\label{sec:error-compensation}

Despite the promising angle accuracy within \(\pm\)\ang{45} around the center, there are apparent deficiencies in directly calculating the \ac{AoA} from the \ac{PDoA}. Therefore, this section discusses and evaluates multiple methods to compensate for the observed measurement error. This includes not only the compensation for the non-linearity but also correcting the wrap-around in \ac{PDoA} measurements and front-back ambiguity. 

Initially, a single correction algorithm is considered to generate a compensation curve for the two sources of \ac{AoA} errors and to calibrate the \ac{TWR} distance measurement. Different \ac{ML} approaches are investigated to model the antenna behavior at different wave incident angles and multi-paths in \ac{LOS} environments. A similar approach has been used in the past for the sole \ac{TWR} distance estimation setup, for example, to extract the \ac{AoA} from a single antenna~\cite{ledergerber2019angle} or to detect \ac{NLOS} conditions~\cite{stahlke2020nlos}.

The dataset used for training and assessing different \ac{ML} models is the same as described in \autoref{sec:empirical-characterization}. It contains 19880 samples (i.e. \ac{TWR} exchanges), of which 19508 were used for training and evaluation while the others where excluded for the following reasons:
\begin{enumerate*}[label=(\roman*),,font=\itshape]
\item \ac{TDoA} values above \qty{1}{\nano\second};
\item \ac{TWR} values below 1000;
\end{enumerate*}
Both of which indicate measurement errors. The remaining samples were split into 11704 samples for training ("train" dataset) and 7804 samples for model evaluation ("test" dataset). In addition, a secondary dataset ("evaluation" dataset) containing 6651 samples was generated using the same data collection setup but with different environmental conditions, such as different room dimensions, furniture, and exact placement of \ac{UWB} modules. This evaluation dataset was used to check the models for generalization into different environments.

The error correction algorithms discussed in this work are restricted by computational load and memory size. The evaluated algorithms should finally work in real-time on a host \ac{MCU}, which typically operates at a frequency of \(\approx\)\qty{100}{\mega\hertz} and provides \(\approx\)\qty{100}{\kilo\byte} of memory. Therefore, the proposed localization system is designed to work independently of the final application and without the support of external computing units, except the host \ac{MCU}.

\subsection{Models and Data Preprocessing}

Two \ac{ML} methods are investigated, namely gradient boosting (using XGBoost~\cite{Chen:2016:XST:2939672.2939785}) and \aclp{NN} (\cite{stahlke2020nlos,ledergerber2019angle,naseri2022machine}). Many different model architectures are found in the literature, therefore different hyper-parameters were evaluated for each model type, always using the same input features and outputs. Model inputs include \ac{PDoA}, \ac{TDoA} and \ac{CIR} measurements at both antennas, the \ac{TWR} distance estimate, and the fraction of the total divided by first path power indicators. The \ac{TDoA} could potentially include information to correct the \ac{PDoA} wrap-around. In contrast, the \acp{CIR} and power indicators potentially include information about the environment, signal quality, and multi-paths, which can support the identification and compensation of antenna non-idealities, including front-back ambiguity. The DW3220 IC provides the \ac{CIR} as 512 complex samples with 36-bit precision each. A subset of these samples was split into real and imaginary parts to be used as \ac{ML} input parameters.

The first set of models is trained for regression, outputting predictions of the \ac{AoA} estimation error and the actual distance between both \ac{UWB} modules. Hyper-parameters evaluated for XGBoost models include the number of estimators between 25 to 500. For neural networks, the following hyper-parameters are evaluated: 
\begin{enumerate*}[label=(\roman*),,font=\itshape]
\item the number of layers (1 to 3);
\item the choice of fully connected layers (32 to 256 nodes);
\item the usage of a first convolutional layer (using filter sizes from 8 to 128 and kernel sizes from 3 to 11).
\end{enumerate*}
Additional settings are evaluated during the training, such as average pooling located after the convolutional layer, or dropout layers with varying dropout rates (0.125 to 0.5), and skip connections (ResNet~\cite{stahlke2020nlos}).

The second set of models is trained for classification of wave incident zones. They are defined as \ang{90} wide \ac{AoA} sections centered around \ang{0}, \ang{90}, \ang{180}, and \ang{270}, hence exactly covering the full circle. The classification is trained and assessed on XGBoost and fully connected neural networks with 1 to 3 layers of 32 to 256 nodes each. Optional dropout layers are re-evaluated as well. Instead of applying for direct compensations, this approach is designed to classify different working zones allowing for subsequent correction of \ac{AoA} and distance measurements with polynomial fitting curves or completely discarding the acquired sample, e.g., in the case of the \ang{180} zone.

\ac{TDoA} and \ac{PDoA} values are scaled to zero mean and unit variance, while the \ac{TWR} distance estimate was scaled to [0,~1]. Of the 512 \ac{CIR} samples from the DW3220, only those between 5 samples before and 100 samples after the first signal path were used. Initial trials with the full \ac{CIR} were computed but did not lead to any tangible improvement; therefore the \ac{CIR} is restricted to decrease the model size and to counteract potential overfitting. The \ac{CIR} values are scaled to a maximum of 1. As target variables for XGBoost, the \ac{AoA} estimation error and the true distance were used. In contrast, the results for \ac{NN}-based models improved by using sine and cosine of the \ac{AoA} estimation error and the true distance as targets.

\subsection{Results}

Both gradient boosting and \ac{NN} models were trained with all possible combinations of the above-mentioned hyper-parameters. The two models proposed here were then selected as a trade-off between the simplicity of the model (for computational efficiency on embedded applications), prediction accuracy on train and test sets, and prediction accuracy on the secondary evaluation dataset.

\begin{figure}[t]
\centerline{\includegraphics[width=\linewidth]{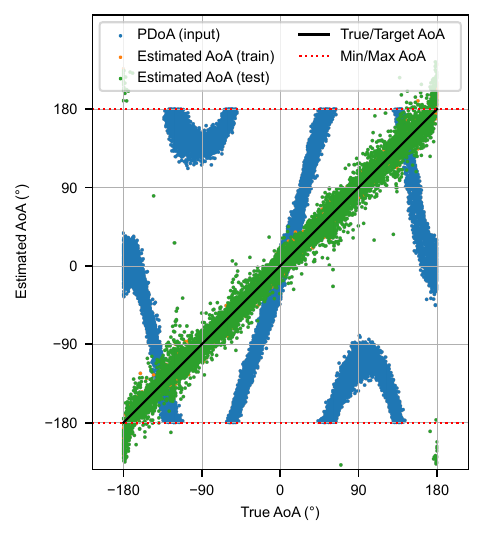}}
\caption{\ac{AoA} estimated by a fully connected \ac{NN} model. The \ac{PDoA} shows the wrap-around and front-back ambiguity, which is both corrected for by the \ac{ML} model. Around \(\pm\ang{180}\), the variance increases, and some estimates lie outside the possible range of \ang{-180} to \ang{180} degrees.}
\label{fig:aoa-correction-ml}
\end{figure}

Using gradient boosting, the chosen model contains 100 estimators resulting in a model size of \qty{42}{\kilo\byte}. The model predictions achieve an \ac{MSE} of \ang{3.21} and \ang{43.70}, respectively, on train and test datasets, over the entire \ang{360} range, effectively solving the front-back ambiguity and \ac{PDoA} wrap-around. The chosen \ac{NN} model consists of a single-layer fully connected \ac{NN} with 224 nodes and a size of \qty{373}{\kilo\byte}. This results in an \ac{MSE} of \ang{2.03} and \ang{19.35} on train and test datasets, respectively. As can be seen in \autoref{fig:aoa-correction-ml}, and as expected, the best accuracy is achieved close to \ang{0}, and accuracy is reduced towards \ang{180}, where significant variance can be observed in the results of the test dataset. Additional improvements can be made by discarding results outside the \ang{-180} to \ang{180} range, which should never occur. This allows for reaching an \ac{MSE} of \ang{1.52} and \ang{15.27} on train and test datasets, respectively.

In addition to the \ac{AoA}, the models corrected the \ac{TWR} distance estimate. XGBoost reached an \ac{MSE} of \qty{0.37}{\centi\meter} and \qty{6.25}{\centi\meter} on the train and test datasets, respectively. Using the single-layer neural network, an improved \ac{MSE} of \qty{0.29}{\centi\meter} on the train dataset and \qty{2.24}{\centi\meter} on the test dataset has been achieved. To ensure the stability over distance (cf. \autoref{sec:empirical-stability}) the \ac{NN} model was again trained on each distance separately. Thereby achieving an \ac{MSE} of between 5.17 and 19.65 for each distance contained in the dataset.

The two models were also assessed on the secondary evaluation dataset to check for possible generalization. This never reached comparable results with the test dataset. Variations in the number of input feature, dropout layers, or skip connections (residual networks) were investigated to reduce overfitting. However, no significant improvement has been noticed. Hence results demonstrate that this methodology is effective for applications where the environment is known \textit{a-priori} or in use cases where a model re-tuning is possible. However, this approach cannot be used for designing plug-and-play and flexible \ac{UWB} sensor nodes that can operate in several indoor/outdoor environments. Moreover, these findings demonstrate how the models effectively learn to compensate for environmental effects, such as reflections, multi-path, and noise, instead of learning the antenna's physical behavior. 

Classification into \ang{90} measurement zones resulted in higher accuracy and better generalization than the regression approach. The XGB classifier achieved perfect precision and recall on the train dataset for all classes. For the center class, precision and recall of 0.98 and 0.98 are achieved on the test dataset and 0.65 and 0.80, respectively, on the evaluation dataset. The fully connected neural network also achieved perfect precision and recall on the train dataset, while on test and evaluation datasets, precision and recall were 0.98, 0.97, and 0.62, 0.87, respectively. This shows the potential of a two-step solution to \ac{AoA} estimation in changing environments. After estimating the zone, a different method can be used to compute the \ac{AoA} estimation (e.g. direct estimation with \autoref{eq:aoa} in the center class).

Finally, regression was performed within the \ang{90} measurement zones. Restricted to one of the three zones not containing \ang{180}, the neural network provided estimations with \ac{MSE} below \ang{0.8} on train and test datasets and below \ang{8} on the evaluation dataset. On the zone containing \ang{180}, an \ac{MSE} of \ang{4.86}, \ang{66.62}, and \ang{212.96} was achieved on train, test, and evaluation datasets, respectively. This shows how restricting the operating \ac{AoA} range can significantly improve the accuracy and generalization of the models. As shown in \autoref{sec:empirical-pdoa-and-aoa}, direct computation of the \ac{AoA} using \autoref{eq:aoa} provides similar results in the center zone. However, the zones at \ang{90} and \ang{270} cannot be computed with the direct approach due to non-linearities and \ac{PDoA} wrap-around.

\section{Conclusion}

This paper introduced a novel dual antenna sensor node using the DW3220 \ac{IC}, thus providing a compact and low-power module for \ac{PDoA} measurements and \ac{AoA}-enabled applications. The node was subsequently used to thoroughly evaluate and characterize the \ac{UWB} \ac{IC} for \ac{AoA} estimations under real operating conditions. Finally, error sources in the \ac{AoA} estimation were identified, and solutions based on different \ac{ML} models were proposed and evaluated.

Instead of relying on connected and tightly synchronized single antenna modules (i.e. based on the DW1000 \ac{IC}) for \ac{PDoA} acquisition, the novel sensor node provides a way to utilize the dual antenna capabilities of the DW3220 \ac{IC}. This not only simplifies the hardware design and decreases \ac{BOM} cost but also reduces the number of error sources. As shown, the remaining errors can often be removed using simple and efficient techniques.

The empirical characterization of the DW3220 \ac{IC} in \autoref{sec:empirical-characterization} showed high \ac{AoA} accuracy within the \ang{90} \ac{AoA} range of \ang{-45} to \ang{45}. With an \ac{MSE} of below \ang{2.41}, the claimed accuracy of the \ac{IC} of \(\pm\ang{5}\) is exceeded. However, the naive approach of computing \ac{AoA} from \ac{PDoA} quickly breaks down if the range of tested angles increases. This stems from non-linearities in the measurement, front-back ambiguity, as well as hardware details, as simulated in \autoref{sec:3DFEM}.

Multiple \ac{ML} models were evaluated, resulting in the detailed analysis of one based on gradient boosting and one based on neural networks. The models are small in size (\qty{21}{\kilo\byte} and \qty{373}{\kilo\byte} for XGB and \ac{NN}, respectively) and will enable integration on low-power hardware. They allow accurate estimation of the \ac{AoA} and distance if trained in the corresponding environment. A physical approach for removing front-back ambiguity is to add a third antenna; however, this might not be necessary as the proposed models could directly correct for that while keeping the \ac{MSE} well below \ang{20}.

Finally, a two-step solution is proposed if the models cannot be trained in the final environment. First, classification into \ang{90} wide measurement zones and second, \ac{AoA} estimation within those zones. The second step can, for example, be computed directly from \ac{PDoA} measurements (\autoref{eq:aoa}) in the zones centered on \ang{0} and \ang{180} \ac{AoA}, or a specific \ac{ML} model for each zone can be used. In addition, error-prone measurements in the zone centered on \ang{180} \ac{AoA} could be discarded.

Possible further research could evaluate the generalizability of \ac{ML} models using an extended dataset or combine multiple double antenna modules for \ac{AoA} estimation and localization in a \ac{3D} space.


\bibliographystyle{IEEEtran}
\bibliography{bstctl,bibliography}

\begin{IEEEbiography}[{\includegraphics[width=1in,height=1.25in,clip,keepaspectratio]{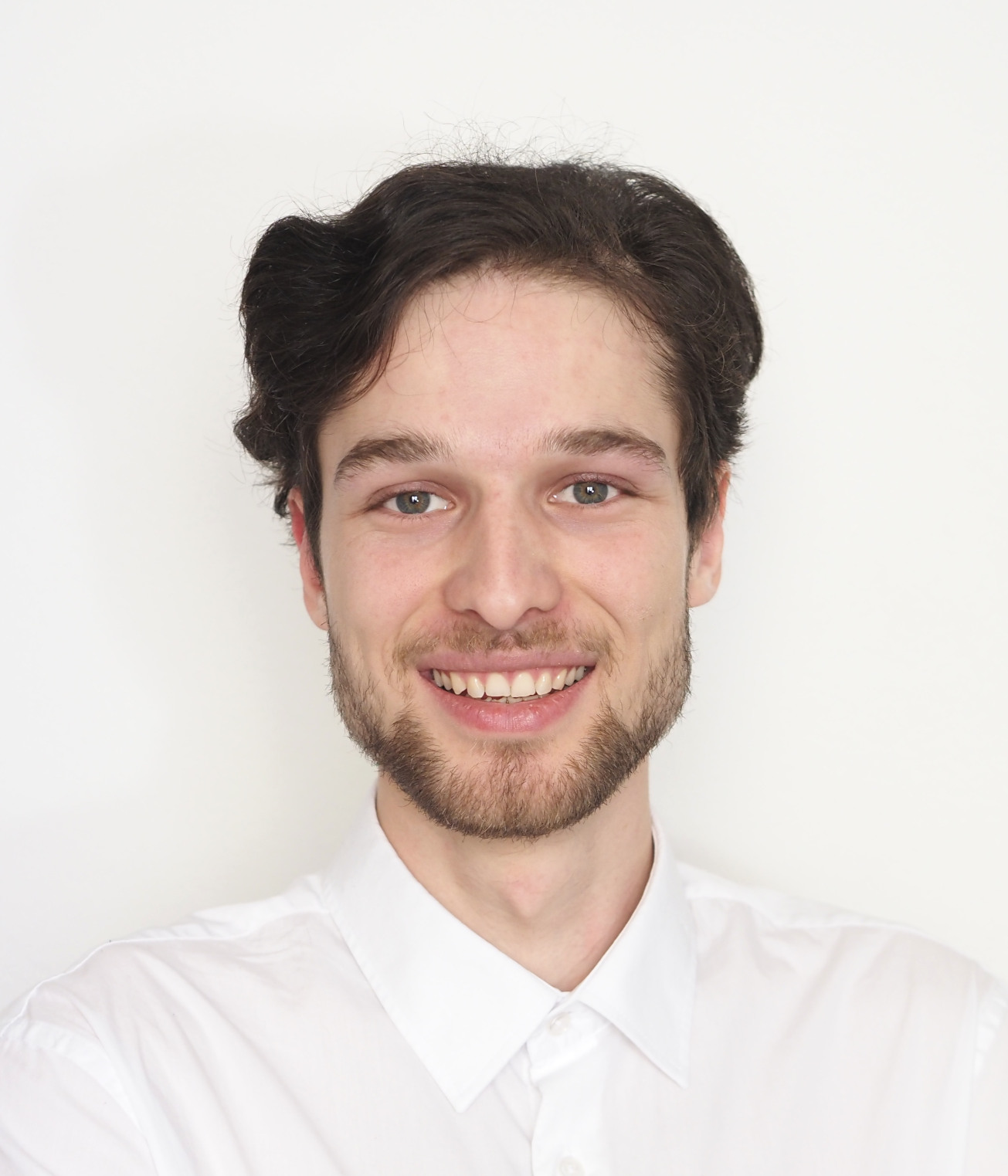}}]{Tobias Margiani} received his B.Sc and M.Sc degree in electrical engineering information technology from ETH Zürich, Zürich, Switzerland in 2021 and 2023, respectively. During his masters studies he spent one semester on exchange at the University of Texas at Austin, Austin TX, USA. His studies focused on embedded systems and communication technology, including indoor localization, UWB, IoT and fault tolerant systems.
\end{IEEEbiography}

\begin{IEEEbiography}[{\includegraphics[width=1in,height=1.25in,clip,keepaspectratio]{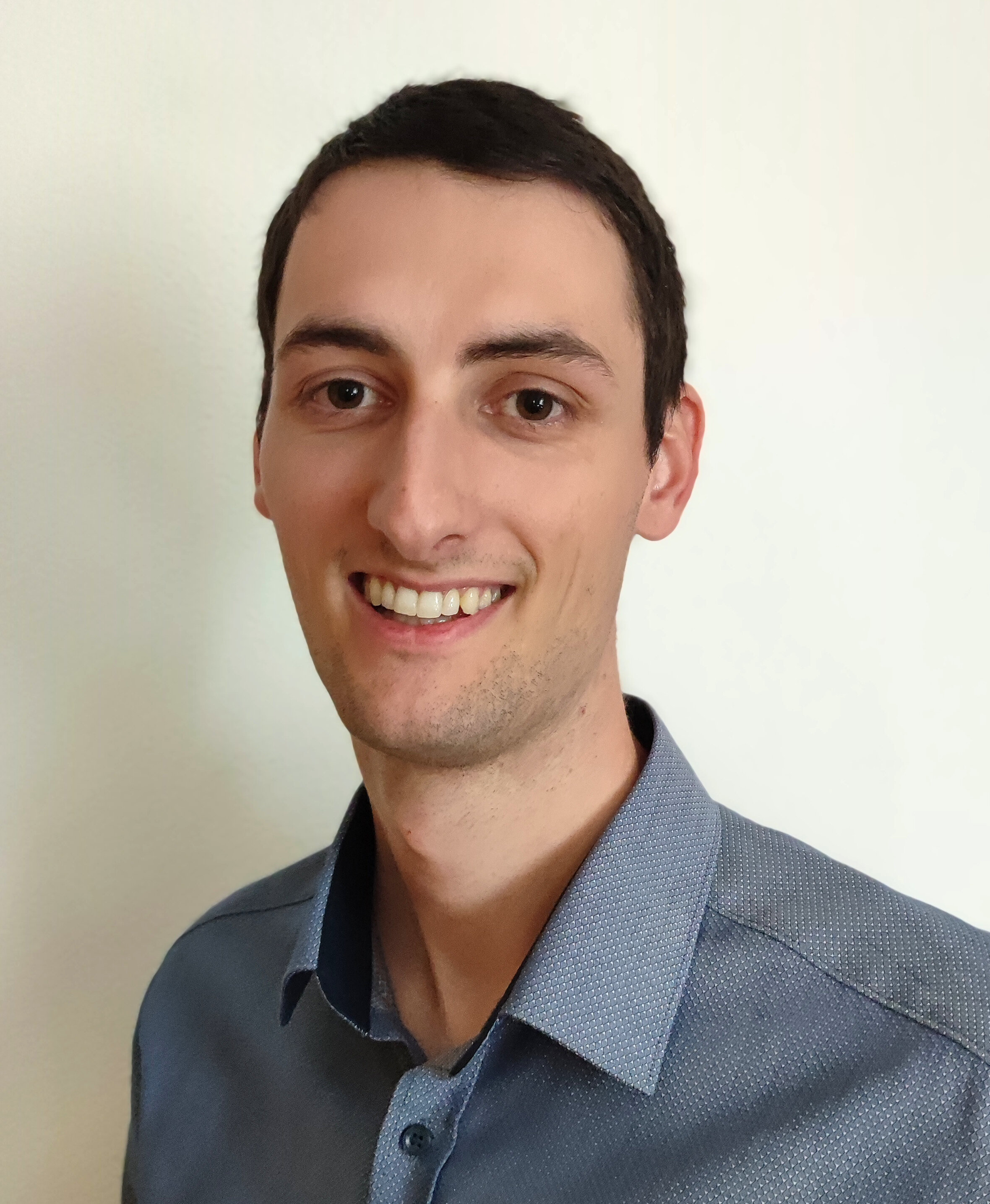}}]{Silvano Cortesi} (S'22) received the B. Sc. and the M.Sc. degree in electronics engineering and information technology from ETH Zürich, Zürich, Switzerland in 2020 and 2021, respectively. He is currently pursuing the Ph.D. degree with the Center for Project-Based Learning at ETH Zürich, Zürich, Switzerland. His research work focuses on indoor localization, ultra-low power and self-sustainable IoT, wireless sensor networks and energy harvesting.
\end{IEEEbiography}

\begin{IEEEbiography}
[{\includegraphics[width=1in,height=1.25in,clip,keepaspectratio]{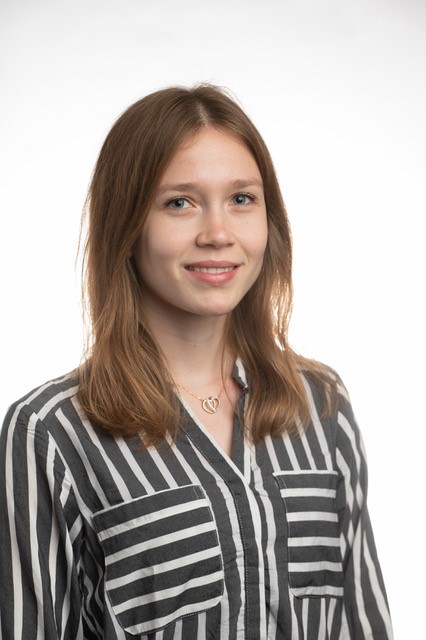}}]{Milena Keller} received the B.Sc and M.Sc degree in electrical engineering from ETH Zürich, Zürich, Switzerland in 2021 and 2023, respectively. She has experience in hardware design for high-frequency radio communication and a good expertise with the most recent indoor localization systems, such as UWB and trilateration. Moreover, she is gaining experience in wireless power transfer and magnetic levitation.
\end{IEEEbiography}

\begin{IEEEbiography}[{\includegraphics[width=1in,height=1.25in,clip,keepaspectratio]{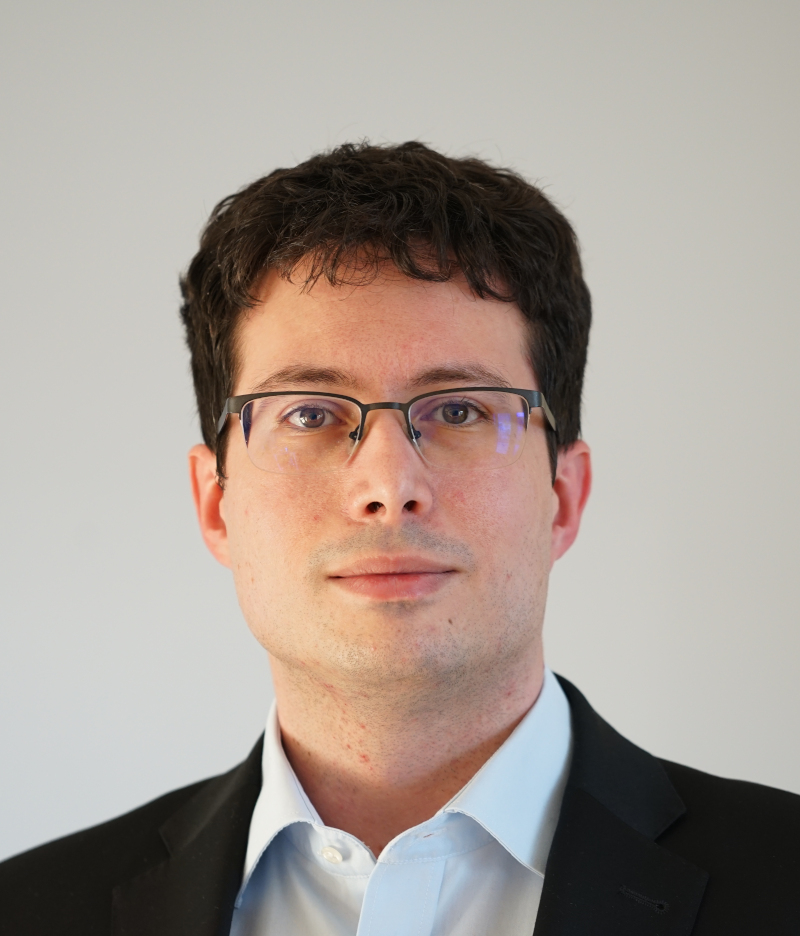}}]{Christian Vogt} (M'20) received the M.Sc. degree and the Ph.D. in electrical engineering and information technology from ETH Zürich, Zürich, Switzerland, in 2013 and 2017, respectively. He is currently a post-doctoral researcher and lecturer at ETH Zürich, Zürich, Switzerland. His research work focuses on signal processing for low power applications, including field programmable gate arrays (FPGAs), IoT, wearables and autonomous unmanned vehicles.
\end{IEEEbiography}

\begin{IEEEbiography}[{\includegraphics[width=1in,height=1.25in,clip,keepaspectratio]{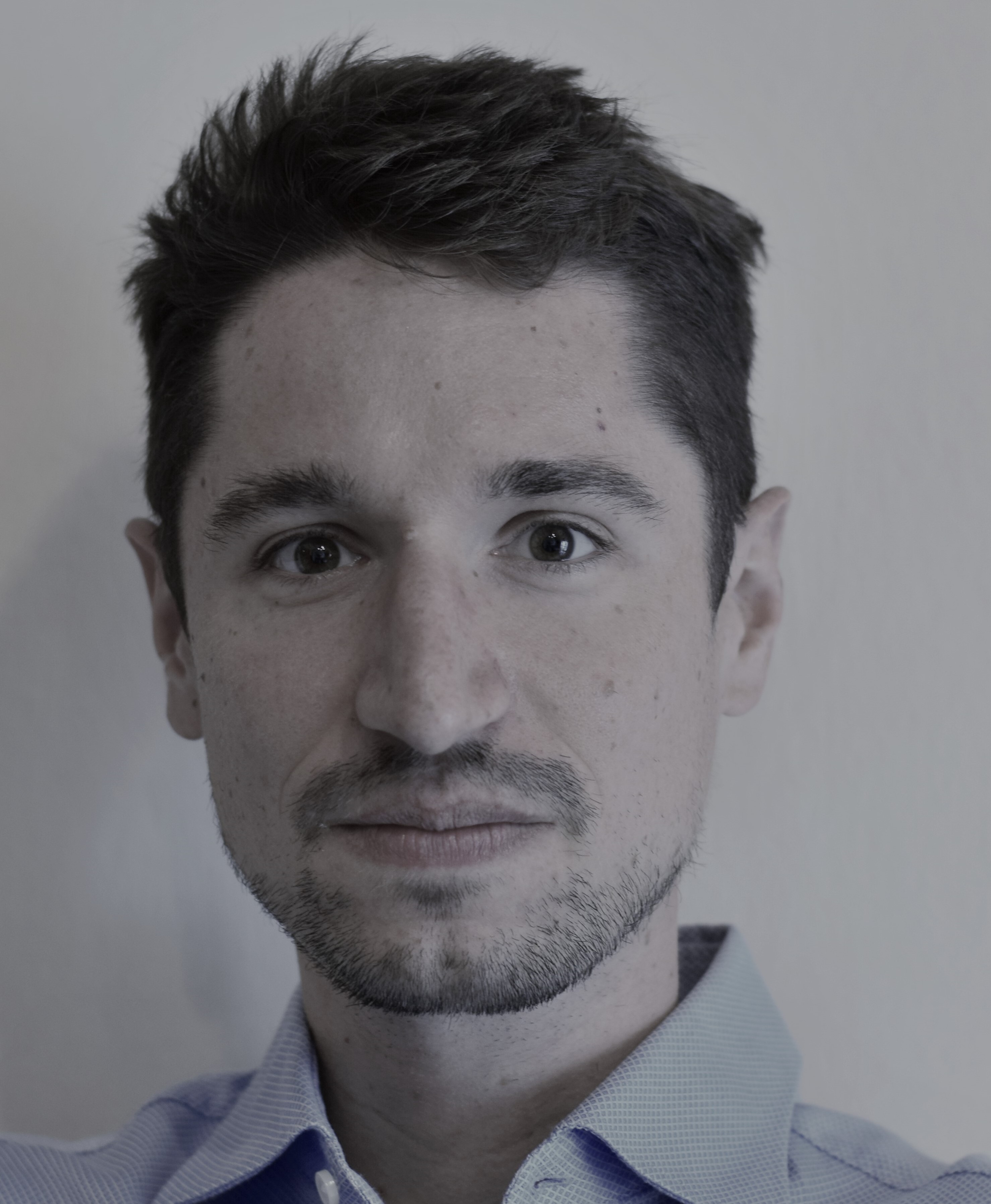}}]{Tommaso Polonelli} (M'20) received the M.Sc. degree and the Ph.D. in electronics engineering from the University of Bologna, Bologna, Italy, in 2017 and 2020, respectively. He is currently a post-doctoral researcher at ETH Zürich, Zürich, Switzerland. His research work focuses on wireless sensor networks, IoT, autonomous unmanned vehicles, power management techniques, structural health monitoring, and the design of ultra-low power battery-supplied devices with onboard intelligence. He has collaborated with several universities and research centers, such as the University College Cork, Cork, Ireland, and the Imperial College London, London, U.K. He has authored over 40 papers in international journals and conferences.
\end{IEEEbiography}

\begin{IEEEbiography}[{\includegraphics[width=1in,height=1.25in,clip,keepaspectratio]{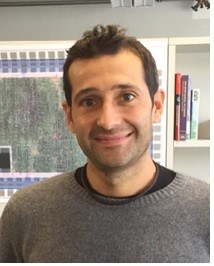}}]{Michele Magno} (SM'13) is currently a Senior Scientist at ETH Zürich, Switzerland, at the Department of Information Technology and Electrical Engineering (D-ITET). Since 2020, he is leading the D-ITET center for project-based learning at ETH. He received his master's and Ph.D. degrees in electronic engineering from the University of Bologna, Italy, in 2004 and 2010, respectively. He is working in ETH since 2013 and has become a visiting lecturer or professor at several universities, namely the University of Nice Sophia, France,  Enssat Lannion, France, Univerisity of Bologna and Mid University Sweden, where currently is a full visiting professor at the electrical engineering department. His current research interests include smart sensing, low-power machine learning, wireless sensor networks, wearable devices, energy harvesting, low-power management techniques, and extension of the lifetime of batteries-operating devices. He has authored more than 220 papers in international journals and conferences. He is a senior IEEE member and an ACM member. Some of his publications were awarded as best papers awards at IEEE conferences. He also received awards for industrial projects or patents.
\end{IEEEbiography}

\end{document}